\documentclass[]{aastex631}

\begin{document}

\title{Numerical simulations of cold clumps in the hot accretion flows around black holes\footnote{Released on March, 1st, 2021}}

\correspondingauthor{Ren-Yi Ma, De-Fu Bu}
\email{ ryma@xmu.edu.cn, dfbu@shnu.edu.cn}

\author{Na-Duo Liu}
\affiliation{Department of Astronomy and Institute of Theoretical Physics and Astrophysics Xiamen University Xiamen, Fujian 361005, China}

\author{Yu-Heng Sheng}
\affiliation{Shanghai Astronomical
Observatory, Chinese Academy of Sciences, 80 Nandan Road, Shanghai 200030, China}

\author{De-Fu Bu}
\affiliation{Shanghai Key Lab for Astrophysics, Shanghai Normal University, 100 Guilin Road, Shanghai 200234, China}

\author{Xiao-Hong Yang}
\affiliation{Department of Physics, Chongqing University, Chongqing 400044, China}

\author{Mao-Chun Wu}
\affiliation{Shanghai Astronomical
Observatory, Chinese Academy of Sciences, 80 Nandan Road, Shanghai 200030, China}
\affil{SHAO-XMU Joint Center for Astrophysics,  Xiamen, Fujian 361005, China}

\author{Ren-Yi Ma}
\affiliation{Department of Astronomy and Institute of Theoretical Physics and Astrophysics Xiamen University Xiamen, Fujian 361005, China}
\affil{SHAO-XMU Joint Center for Astrophysics,  Xiamen, Fujian 361005, China}

\begin{abstract}
Previous numerical simulations have shown that cold clumps can form within hot accretion flows, offering insights into the detailed processes of the state transition in black hole X-ray binaries.
However, the evolution of the cold clumps has not been investigated in detail yet.
In this paper, we conduct hydrodynamic simulations to investigate the evolution of the cold clumps.
In addition to previous result that when the accretion rate is high enough the cold clumps emerge within the hot accretion flow,
we found that instead of directly moving toward to the black hole, the clumps moves outward when they initially form.
The reason should be the combination of viscous torque and the condensation of hot gas from larger radii, which lead to the slightly super-Keplerian angular momentum of the clumps. 
After reaching the equilibrium position, the clumps begin to fragment at the inner edge with each fragment moving inward sequentially.
Generally, the azimuthal movement of the clumps are quasi-Keplerian, being closer to the outer detached Keplerian cold disk rather than the surrounding sub-Keplerian hot accretion flow, which agrees well with the semi-analytical results for weak coupling case in \cite{Wang2012}. 
\end{abstract}


\section{Introduction} \label{sec:intro}

As is well known, black hole X-ray binaries exhibit different spectral states during outbursts, such as the hard state (HS), soft state (SS) and intermediate state (IMS), which are characterized by their X-ray spectral and timing properties \citep[e.g.,][]{Zdziarski2004,2006ARA&A..44...49R,2010LNP...794...53B}. 

The theoretical models of the accretion flow for SS and HS have been widely accepted. The former can be well explained with the standard accretion disk model, or the cold disk, in which the thermal emissions of the cold disk dominates the X-ray spectra\citep[e.g.,][]{1973A&A....24..337S,1998bhad.book.....K}.
For the latter state, the accretion flow is believed to be the hot accretion flow that truncates the outer cold disk. It is hot because the released gravitational potential energy is mainly stored in the gas and advected into the central black hole. Its comptonization on the seed photons from synchrotron radiation by itself or thermal emission by the cold disk outside the truncation radius contributes the hard power law component of the spectra \citep[e.g.,][]{1994ApJ...428...13N, YN2014}.  

While for IMS, or the transitional state between HS and SS, the accretion mode is still on debate.
Firstly, \citet[][]{Esin1997} proposed that the truncation radius moves inward until it reaches the ISCO during transition from hard to soft state, which means the accretion mode is similar to that of HS but with smaller truncation radius. 
Secondly, considering the balance between the evaporation and condensation, an inner cold debris may form in the hot accretion flow during IMS as the condensing rate is higher in the inner region of the hot accretion flow \citep{Liu2007,Liu2011,Qiao2011,Liu2022}. 
Thirdly, as a result of the thermal instability inherent in the accretion flows, the cold gas within a hot accretion flow may be present in the form of clumps or clouds when the timescale of thermal instability is shorter than the infall timescale \citep{2003ApJ...594L..99Y}. Such clumps have been observed or supported in quite a few simulations \citep{2016MNRAS.459.1543W,2018MNRAS.474.1206B,2021ApJL...919...20D}.
Additionally, the clumps are possible to form an inner debris due to the tidal forces and differential rotation \citep{Wang2012}
Studies have been done to find observational constraints on these scenarios. 
\citet{Yu2018} investigated how the profiles of the broad iron line are affected for different scenarios.
Combined with the observed variation of the iron line during state transition of MAXI J1631-479 \citep{Xu2020},
\citet{Shui2023} investigated the relation between the line width and flux, and found the clump model are favored.
If the geometric variation of the clump can be given, either by theoretical analysis or numerical simulations, it is possible to constrain the scenarios better.

Although previous simulations have identified the formation of clumps, they focused mostly on the process of state transition, without detailed investigation and analysis of the clumps.
Based on the Boltzmann Equation, \citet{Wang2012} investigated the dynamics of clumps embedded in the hot accretion flows. However, the formation of the clumps is assumed. 
\citet{Xie2012} studied the two-phase accretion flow based on global solution, but they concerned mostly on the radiative efficiency of the hot gas.
In this paper, we presented two-dimensional hydrodynamic simulations of the accretion flows around stellar-mass black holes at different accretion rates, focusing on the evolution of cold clumps. 

This paper is organized as follows. In section 2, we introduce our numerical methods. Our results are given in section 3. Finally, we discuss and summarize the results in section 4.

\section{Numerical Method} \label{sec:Method}

\subsection{Basic equations}
We perform two-dimensional axis-symmetric hydrodynamical simulation using the ZEUS-MP code \citep{2006ApJS..165..188H}, in which the spherical coordinates ($R$,$\theta$,$\phi$) are employed. The basic equations of the back hole accretion flow are as follows,

\begin{equation}
  \frac{d \rho}{d t} + \rho \nabla \cdot \mathbf{v} = 0
\end{equation}
\begin{equation}
  \rho \frac{d \mathbf{v}}{d t} = -\nabla P - \rho \nabla \Phi + \nabla \cdot \mathbf{T}
\label{Eq.2}
\end{equation}
\begin{equation}
  \rho\frac{d (e/\rho)}{d t} = - P \nabla \cdot \mathbf{v} + \mathbf{T}^2/\mu + \rho  \mathcal{L},
\end{equation}

\noindent where \(\rho\), $\mathbf{v}$, \(e\), \(P\) are the density, velocity, internal energy, and pressure of gas, respectively. \(\mathbf{T}\)  and \(\mathcal{L}\) represent the viscous stress tensor and radiative cooling rate per unit mass, respectively. The quantity $\mu \equiv \rho \nu$, with $\nu \sim \alpha c^2_s / \Omega_{\text{K}}$ being the kinematic viscosity, $c_s$ being the sound speed and $\Omega_{\text{K}}$ being the Keplerian angular velocity. 
The ideal gas with \(P = (\gamma - 1)e\) and the adiabatic index \(\gamma = 5/3\) is taken. In this paper, we set the black hole mass to a typical value of $M_{\text{BH}} = 10 M_{\odot}$ with $M_{\odot}$ being the solar mass. The potential is assumed to be pseudo-Newtonian, i.e.,  \(\Phi = -GM_{\text{BH}}/(R-R_S)\), where $R_S \equiv 2GM_{\text{BH}}/c^2$ is the Schwarzschild radius. 
As the azimuthal components of $\mathbf{T}$ dominate other components \citep{Stone2011}, we only consider the azimuthal components of the viscous stress tensor for the convenience of dicussion, following previous works as \citep[e.g.,][]{1999MNRAS.310.1002S,2019ApJ...875..147B}, 

\begin{equation}
\mathbf{T}_{R \phi} = \mu R \frac{\partial}{\partial R} \left( \frac{v_{\phi}}{R} \right),
\end{equation}
\begin{equation}
\mathbf{T}_{\theta \phi} = \mu \sin \theta \frac{\partial}{\partial \theta} \left( \frac{v_{\phi}}{R\sin \theta} \right).
\end{equation}

\noindent Since $c_s$ and $\Omega_{\text{K}}$ are proportional to $R^{-1/2}$ and $R^{-3/2}, respectively$, $\nu = \alpha \sqrt{GM_{\text{BH}}r}$ is taken in our simulations. In this paper, we use $\alpha = 0.01$ as default, and for certain cases $\alpha = 0.02$ is taken for the purpose of comparison.

\subsection{Initial and boundary conditions}
In the radial direction, the simulation region ranges from $1.25 R_S$ to $500 R_S$. For the $\theta$ direction, we have $0<$\(\theta\)$<$ \(\pi\). In the radial direction, the domain is divided into 192 logarithmically spaced grids. In the $\theta$ direction, 88 uniformly spaced grids are adopted.
We use the Keplerian orbital time at $100 R_S$ as the time unit, as denoted by $t_g$. The duration of the simulation ranges from 0 to 10$t_g$. 
On both the outer and inner radial boundaries, the outflow boundary conditions are employed, which prevent gas from flowing into the computational domain. On the boundaries of $\theta = 0$ and $\pi$, the axis-symmetric boundary conditions are taken.

\begin{figure*}
\centering
\includegraphics[width=0.3\linewidth]{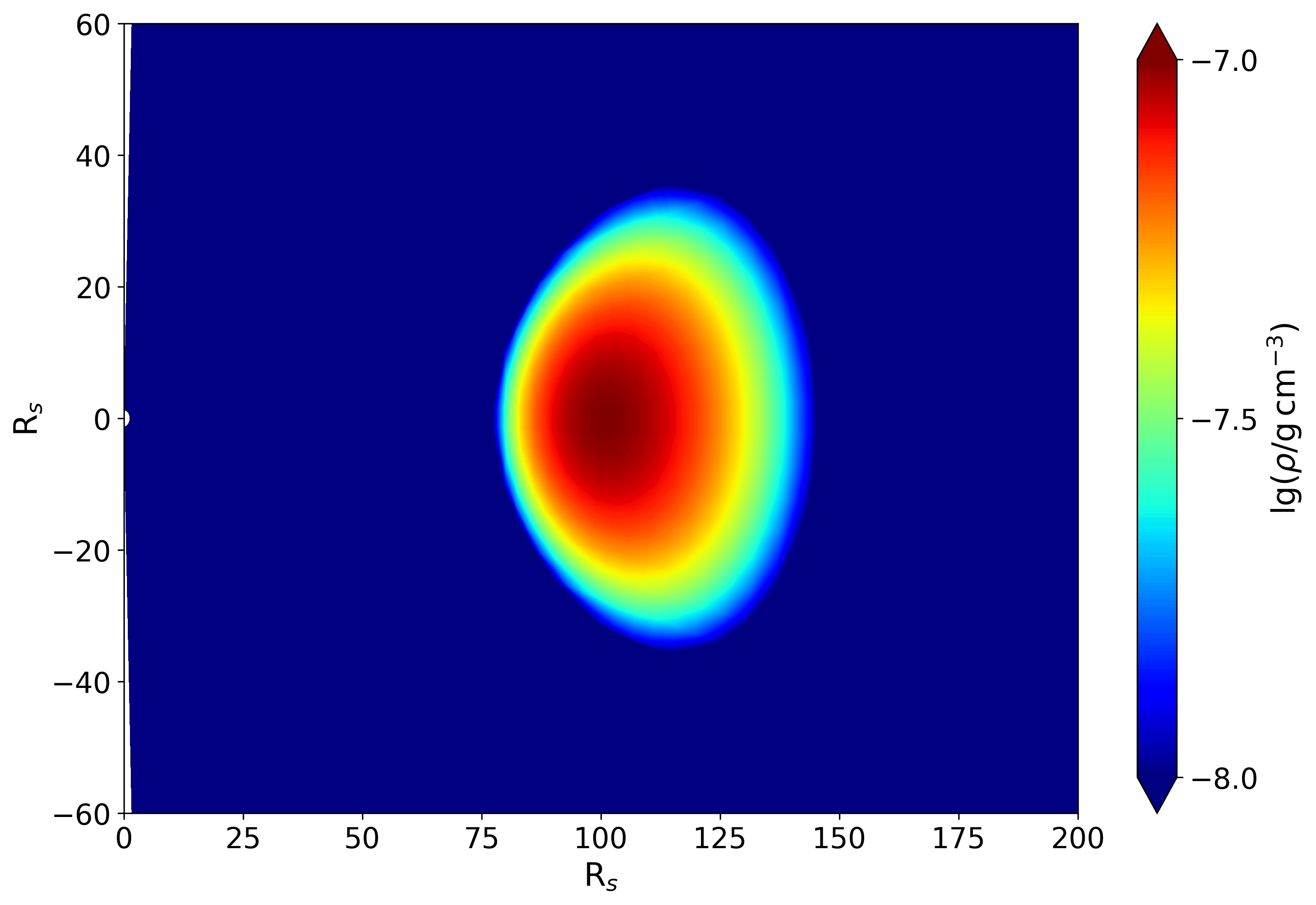}
\includegraphics[width=0.3\linewidth]{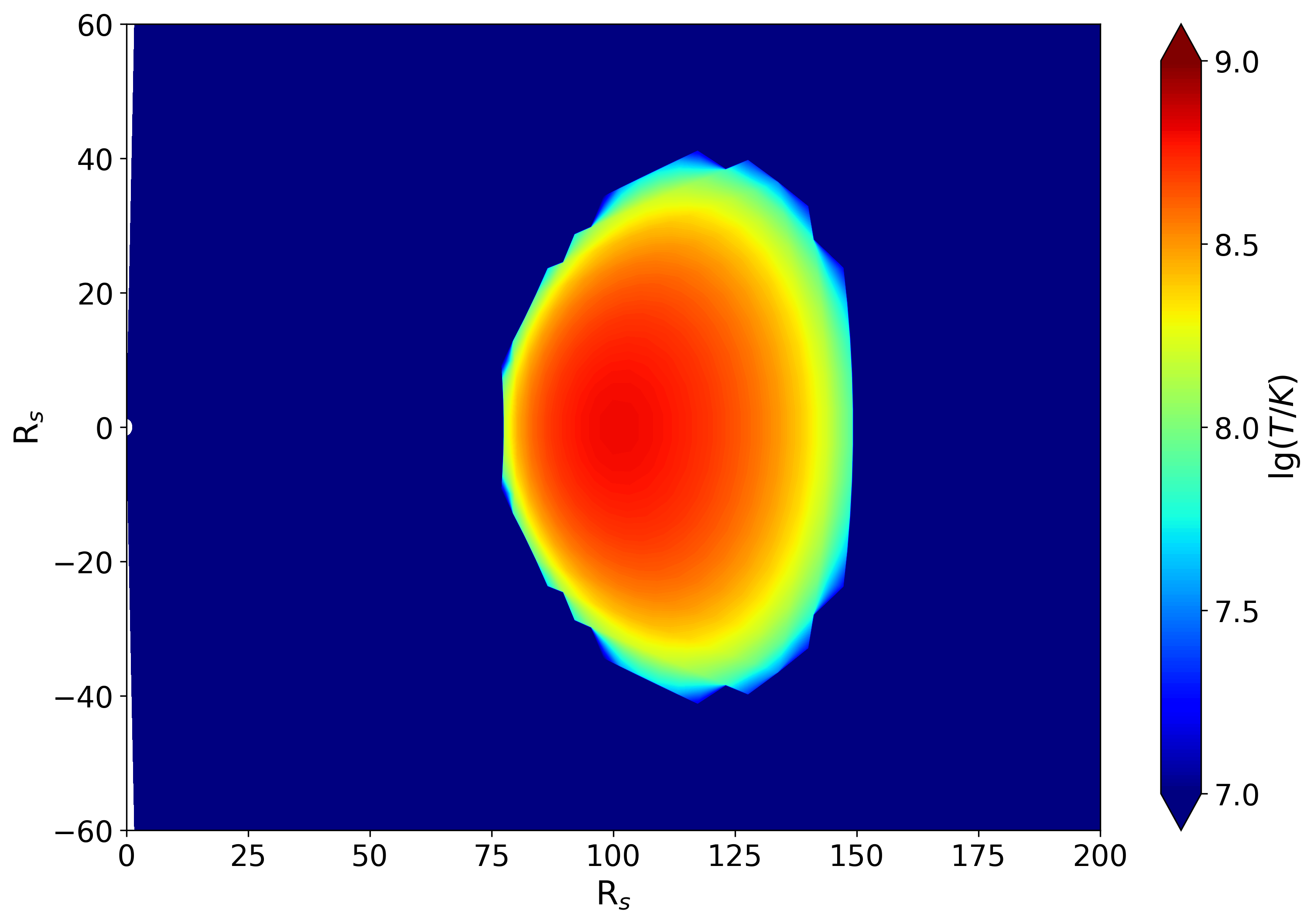}
\includegraphics[width=0.3\linewidth]{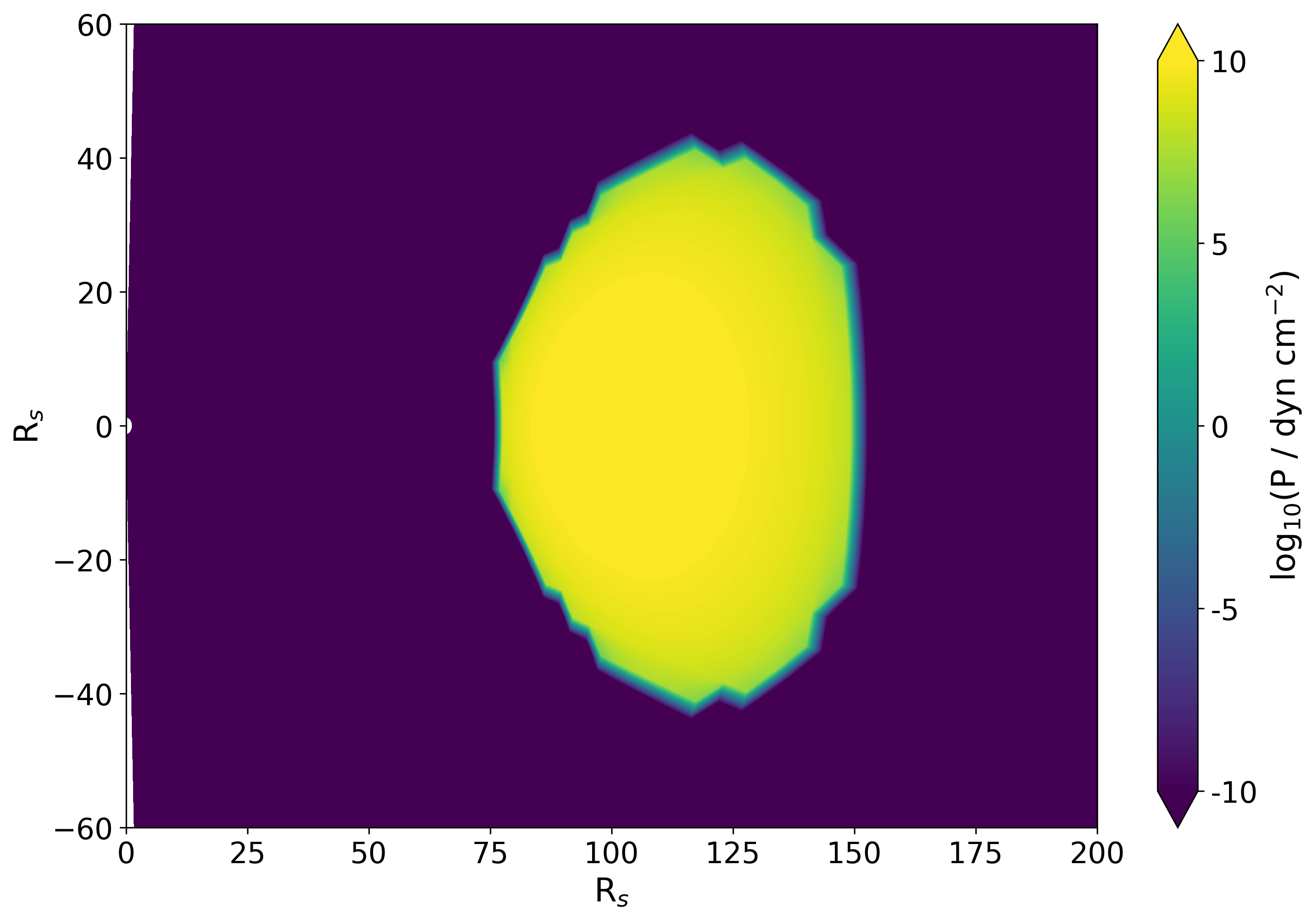}
\caption{Initial conditions of the torus for the case $\rho_{\text{max}} =10^{-7}\text{g/cm}^3$. The {\it left}, {\it central} and {\it right} panels correspond to the initial density, temperature and pressure, respectively.
\label{fig:initial}}
\end{figure*}

The initial condition is a torus rotating with a constant specific angular momentum embedded in very low density medium gas. The pressure and density of the torus are related with a polytropic equation of state $p \propto \rho^\gamma$ and the structure of the torus is given by \citep[e.g.,][]{1984MNRAS.208..721P}
\begin{equation}
\frac{p}{\rho} = \frac{(\gamma-1)GM}{\gamma{R_0}} \left[ \frac{R_0}{R} - \frac{1}{2} \left( \frac{R_0}{R \sin \theta} \right) - \frac{1}{2d} \right],
\end{equation}
\noindent where $R_0$ represents the location of the torus center, the density of which is maximal and denoted by $\rho_{\text{max}}$.  The parameter of $d$ describes the distortion of the equilibrium torus. 
Here we take $R_0 =100 R_S$ and $d$ = 1.125. The inner and outer boundaries of the torus are located at 75$R_S$ and 150$R_S$. 
The initial torus is embedded in a low-density background gas. The
density of background gas should be low enough to guarantee that at
the torus surface there are no artificial shocks or oscillations induced by artificial pressure gradient\citep{2003MNRAS.344..978R, 2003MNRAS.344L..37R,2006MNRAS.369.1235B,2017MNRAS.467.4036M}. Following \citet{2018MNRAS.474.1206B}, we set the background density to be four orders of magnitude smaller than the torus edge density, and the background temperature is $10^{4}\,\mathrm{K} $.
Figure~\ref{fig:initial} shows the initial density, temperature and pressure profiles of the torus.

\begin{figure}
\centering
\includegraphics[width=0.45\linewidth]{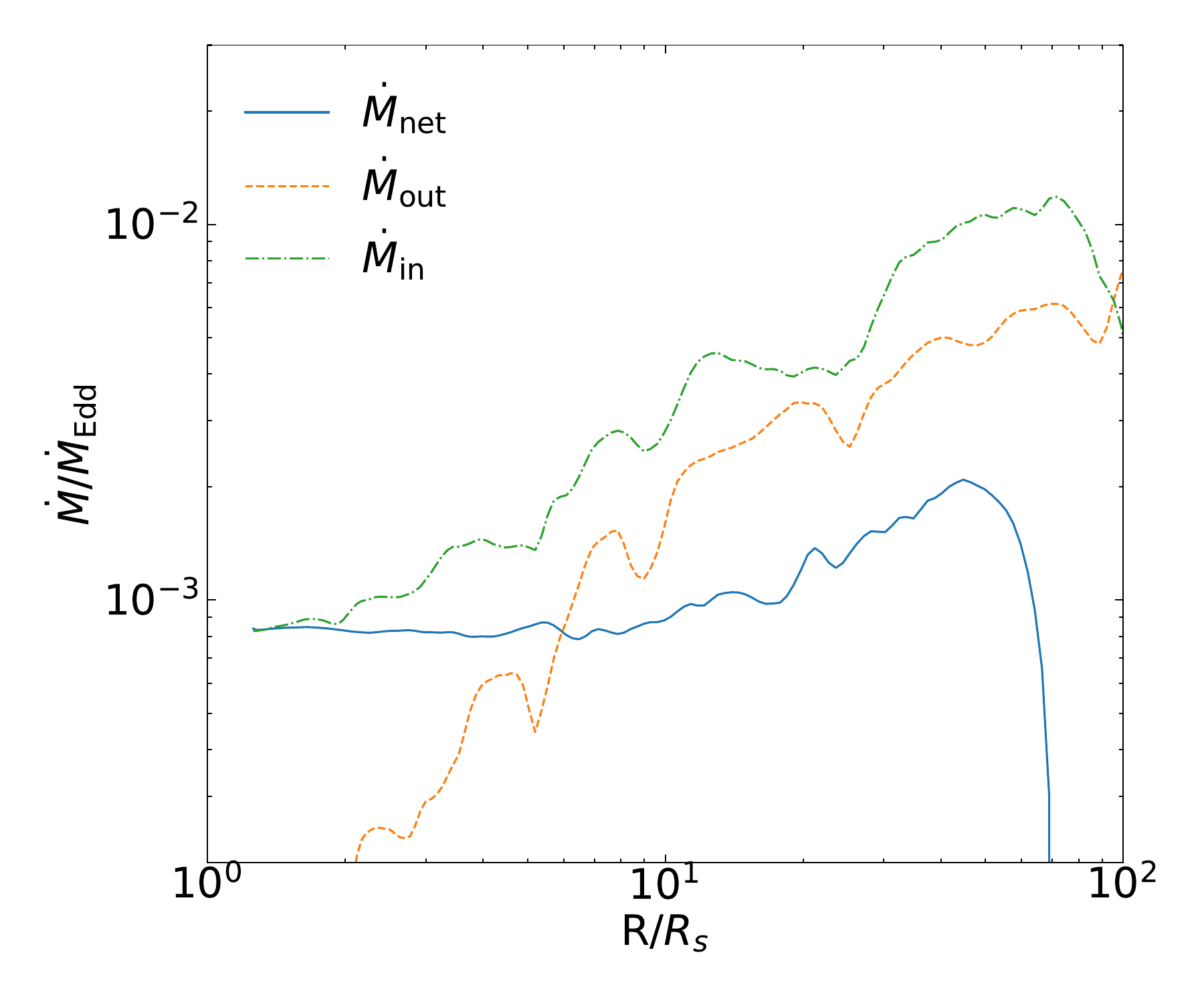}
\caption{\label{fig:C5}Time-averaged (from 0.8~$t_g$ to 1.0~$t_g$) radial distributions of mass fluxes of model C3.}
\end{figure}

We first run the simulations without radiative cooling to a quasi-steady state, and then turn on the radiative cooling to obtain clumps. Whether the accretion flow becomes steady can be checked with the net accretion rate, which describes the difference between the inflow rate and outflow rate.
Here the mass inflow rate, mass outflow rate, and the net accretion rate at radius $r$ are as follows,
\citep{1999MNRAS.310.1002S,2019ApJ...875..147B},
\begin{equation}
\dot{M}_{\text{in}}(R) = -2\pi R^2 \int_0^\pi \rho \min(v_r, 0) \sin \theta d\theta,
\end{equation}
\begin{equation}
\dot{M}_{\text{out}}(R) = 2\pi R^2 \int_0^\pi \rho \max(v_r, 0) \sin \theta d\theta,
\end{equation}
\begin{equation}
\dot{M}_{\text{net}}(R) = \dot{M}_{\text{in}}(R) - \dot{M}_{\text{out}}(R).
\end{equation}
For a steady accretion flow, the net accretion rate is a constant function of radius. So when the net accretion rate becomes almost constant, we can roughly identify the accretion flow to be in quasi-steady state.
Figure~\ref{fig:C5} shows the radial profiles of the mass inflow rate $\dot{M}_{\text{in}}$, outflow rate $\dot{M}_{\text{out}}$ and net rate $\dot{M}_{\text{net}}$ in model C3, time-averaged over the period from $0.8 t_g$ to $1.0 t_g$. The net accretion rate is almost a constant function of radius inside $20R_S$, which indicates that the accretion flow has reached a quasi-steady state in this region.

\subsection{Cooling rate and clump criteria}
Various radiation mechanisms, such as synchrotron, inverse Compton scattering and bremsstrahlung, should play roles in the hot accretion flow. However, we consider only the bremsstrahlung radiation for simplicity. Since the main purpose of this paper is to investigate the formation and movement of the clumps instead of the exact critical conditions of transition, such simplification does not affect our results significantly. Following the approach described by \citet{2000ApJ...543..686P}, we describe the cooling rate with the following equation,
\begin{equation}
    \rho \mathcal{L} = - 3.3 n^2 T^{1/2}\times 10^{-27} \quad {\text{erg cm}}^{-3} \text{s}^{-1},
\end{equation}
\noindent where $n$ is the number density and $T$ is the temperature of gas. It should be noted that to avoid over collapsing of cold clouds, as well as to reduce the complexity of simulation when the clumps become optically thick, only the radiative cooling during the formation of clumps are considered. Once the clumps form, we turn off the radiative cooling in the clumps region. 
As our results show, although the viscous heating is still there, it does not heat up the clumps significantly. %which  be because the clumps move as a whole to some extend, with the differential rotating being reduced.
%So our simplification to stop the radiative cooling of the clumps does not obviously affect the simulations. 

The way to identify the cold clumps in the hot accretion flow is easy to understand. 
The temperature should be as low as the standard cold disk, i.e. about $10^5\,{\text{K}}$. 
However, some unstable region could also reach such temperature. 
So we set the criterion of temperature to be even lower as $T < 10^{4}\,{\text{K}}$.
Although the background gas is set to be of such temperature, as mentioned above, it is easy to be excluded with extremely low density.
Therefore, in this paper, the clumps are defined with the following criterion,
\begin{equation}
    T < 10^4\, {\text{K}} \quad \& \quad \rho > 10^{-5} {\rho_{\text{max}}}.
\end{equation}

\begin{table}
	\centering
        \small
	\caption{Simulation parameters}
	\label{tab:simulationparams}
	\begin{tabular}{lccccc} % four columns, alignment for each
		\hline
        Model & $\alpha$ &  $\rho_{\text{max}}$ (g/cm$^3$)  & $\mathrm{\dot{M}_{BH}}/\mathrm{\dot{M}_{Edd}}$ & Cooling & $t_\mathrm{rad}$\\
        \hline
        C0 & 0.01 & $5.2 \times 10^{-8}$  & $3.0 \times 10^{-4}$ & off & /\\
        C1 & 0.01 & $3.6 \times 10^{-8}$  & $6.3 \times 10^{-4}$ & on & $1.0t_g$ \\
        C2 & 0.01 & $4.4 \times 10^{-8}$  & $7.3 \times 10^{-4}$ & on & $1.0t_g$ \\
        C3 & 0.01 & $2.8 \times 10^{-8}$  & $7.6 \times 10^{-4}$ & on & $1.0t_g$ \\
        C4 & 0.01 & $6.0 \times 10^{-8}$  & $1.2 \times 10^{-3}$ & on & $1.0t_g$ \\
        C5 & 0.01 & $9.2 \times 10^{-8}$  & $1.2 \times 10^{-3}$ & on & $1.0t_g$ \\
        C6 & 0.01 & $9.6 \times 10^{-8}$  & $1.3 \times 10^{-3}$ & on & $1.0t_g$ \\
        C7 & 0.01 & $1.0 \times 10^{-7}$  & $1.4 \times 10^{-3}$ & on & $1.0t_g$ \\
        C8 & 0.01 & $1.4 \times 10^{-7}$  & $2.1 \times 10^{-3}$ & on & $1.0t_g$ \\
        B1 & 0.02 & $1.0 \times 10^{-7}$  & $4.5 \times 10^{-3}$ & on & $1.0t_g$ \\
        B2 & 0.02 & $3.0 \times 10^{-7}$  & $1.0 \times 10^{-2}$ & on & $1.0t_g$ \\ 
		\hline
	\end{tabular}
\end{table}

Table~\ref{tab:simulationparams} shows a selection of models from our simulations, 
in which the model ID, the value of viscous parameter, the density of the torus center,  the mass accretion rate at the inner boundary when the accretion becomes stable, Cooling Activation, Radiative Cooling Start Time are listed in Column 1-6, respectively. 
Model C0 simulates a stable hot accretion flow without radiative cooling, which is used as a control group.
Models C1-C8 are the same as Model C0, but correspond to different accretion rates, or the initial densities of the torus, which vary in the range $10^{-14}$-$10^{-12}$ g/cm$^3$. 
Models B1 and B2 are used to explore the influence of the viscous parameter, in which $\alpha=0.02$ instead of 0.01.

\section{Results}

\begin{figure*}  
\centering
\includegraphics[width=1\linewidth]{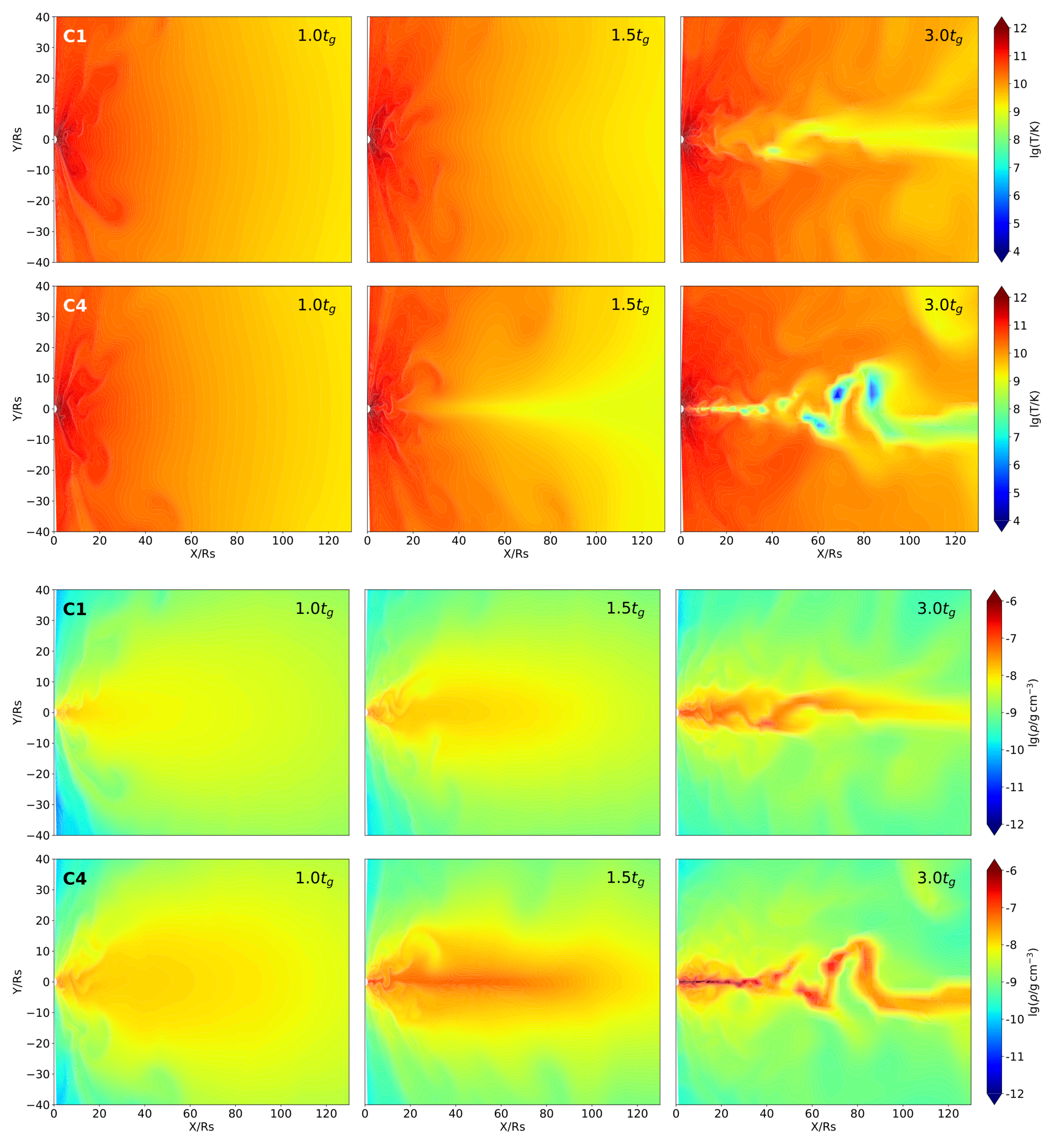}
\caption{\label{fig:DT}The distribution of density and temperature at different time in Model C1 and C4. The upper two rows display the temperature distribution in simulations C1 and C4, respectively. And the lower two rows
show the density distribution. From left to right, the first column shows the time slice at 1.0~$t_g$  before turning on radiative cooling. The second column corresponds to 1.5~$t_g$, and the third column is at 3.0~$t_g$ to show the evolution of accretion after the cooling is turned on. } 
\end{figure*}

\subsection{Mass fraction of the clumps}

In our simulations, accretion rate is the central parameter of interest. For low accretion rate, such as Models C1-C3, the clumps do not form no matter what values the other parameters are. 
While for the cases of high enough accretion rates, such as Models C4-C8, the clumps can form. 
As an example, Figure~\ref{fig:DT} demonstrates the temporal evolution of temperature and density for Models C1 and C4. 
It is clear that although the temperature distributions of the two models are quite similar before turning on radiative cooling ($ t = 1.0t_g $), they show great differences after turning on radiative cooling ( $ t > 1.0t_g $). For Model C4, the temperature decreases more signifcantly and cold clumps finnaly form before the accretion rate reaches maximum, while for Model C1, the reduced temperature is insufficient to form cold clump even when the accretion rate reaches maximum.

\begin{figure*}
\centering
\includegraphics[width=1\linewidth]{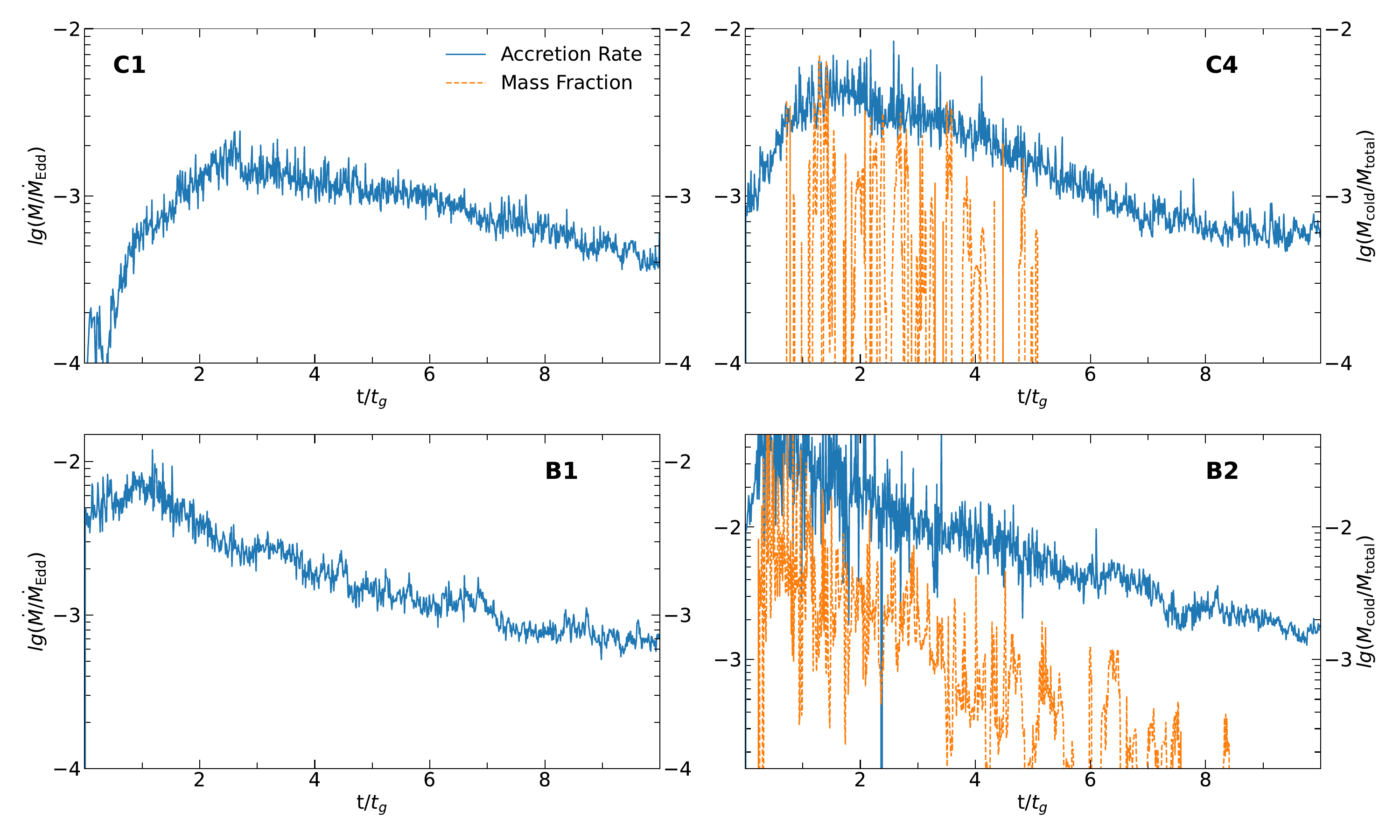}
\caption{\label{fig:line}The time evolution of mass fraction of the cold clumps and the accretion rate in Simulations C1, C4, B1, B2. The right-vertical axis in each panel represents the ratio of cold gas mass to total mass. The left-vertical axis in each panel represents the black hole accretion rate.}
\end{figure*}

Figure~\ref{fig:line} quantatively shows the time evolution of the accretion rate and the mass fraction of cold clumps relative to the entire accretion flow for different models. The blue solid lines show the time evolution of the accretion rate, while the orange dotted lines manifest the presence of cold clumps, as well as the time evolution of clump mass fraction.
The accretion rate first increases to maximum and then decreases gradually due to the internal transfer of angular momentum \citep[e.g.,][]{Igumenshchev1996}. 
Here we mainly concern the mass fraction.
It can be seen that, during the rising stage, the clumps form when the accretion rate exceeds some cirtical value. In the upper left panel, the accretion rate is always lower than ~0.2\% $\dot{M}_{Edd}$, and there exists no clump. While when the accretion rate is higher than  ~0.2\% $\dot{M}_{Edd}$ clumps appear as shown in the upper right panel. 
Moreover, the critical accretion rate is related with the value of $\alpha$. 
As the lower two panels show, for the cases of $\alpha=0.02$, during the stage when the accretion rate keep rising, the critical accretion rate is about ~1.0\% $\dot{M}_{Edd}$, which is higher than the cases of $\alpha=0.01$.
Our results about the critical accretion rate are consistent with previous theoretical and numerical results\citep[e.g.,][]{YN2014, 2016MNRAS.459.1543W}.

During the decaying stage, or when the accretion rate decreases, the existing clumps start to dissipate back into the hot gas, as shown by the decreasing mass fraction. 
As shown in the right panels of Figure~\ref{fig:line}, the clumps remain existence until the accretion rate is as low as  ~0.1\% and 0.2\% $\dot{M}_{Edd}$, respectively, which are lower than the above critical accretion rate during the rising stage.
It is reasonable as the dissipation of clumps takes time, which depends on the mass and dissipation rate of the clumps.
Since the situation of decaying stage is more complex, we concentrate mostly on the rising stage in the following studies.

\subsection{Movement of the clumps}
The radial movement of clumps is complex, especially during the decaying stage, as they may move outward, inward, or even remain stationary for a short period. But for the rising stage, the movement is relatively simple, and we investigated in more details.

Considering the irregular geometry, it is difficult to identify a representative point like the mass center to track the clump position and to study its movement. 
Since we mainly concern the radial movement, meanwhile the clump temperature is significantly lower than the surrounding hot gas, it is feasible to use the radial position of minimal vertically mass weighted averaged temperature to describe the clump center. 
To exclude the hot gas as much as possible, so that the clump mass is comparable to or higher than that of the hot gas, when calculate the averaged temperature we limit the vertical region to be 88°$<\theta<$ 92°, as the clumps predominantly form near the equatorial plane.

\begin{figure*}
\centering
\includegraphics[width=1\linewidth]{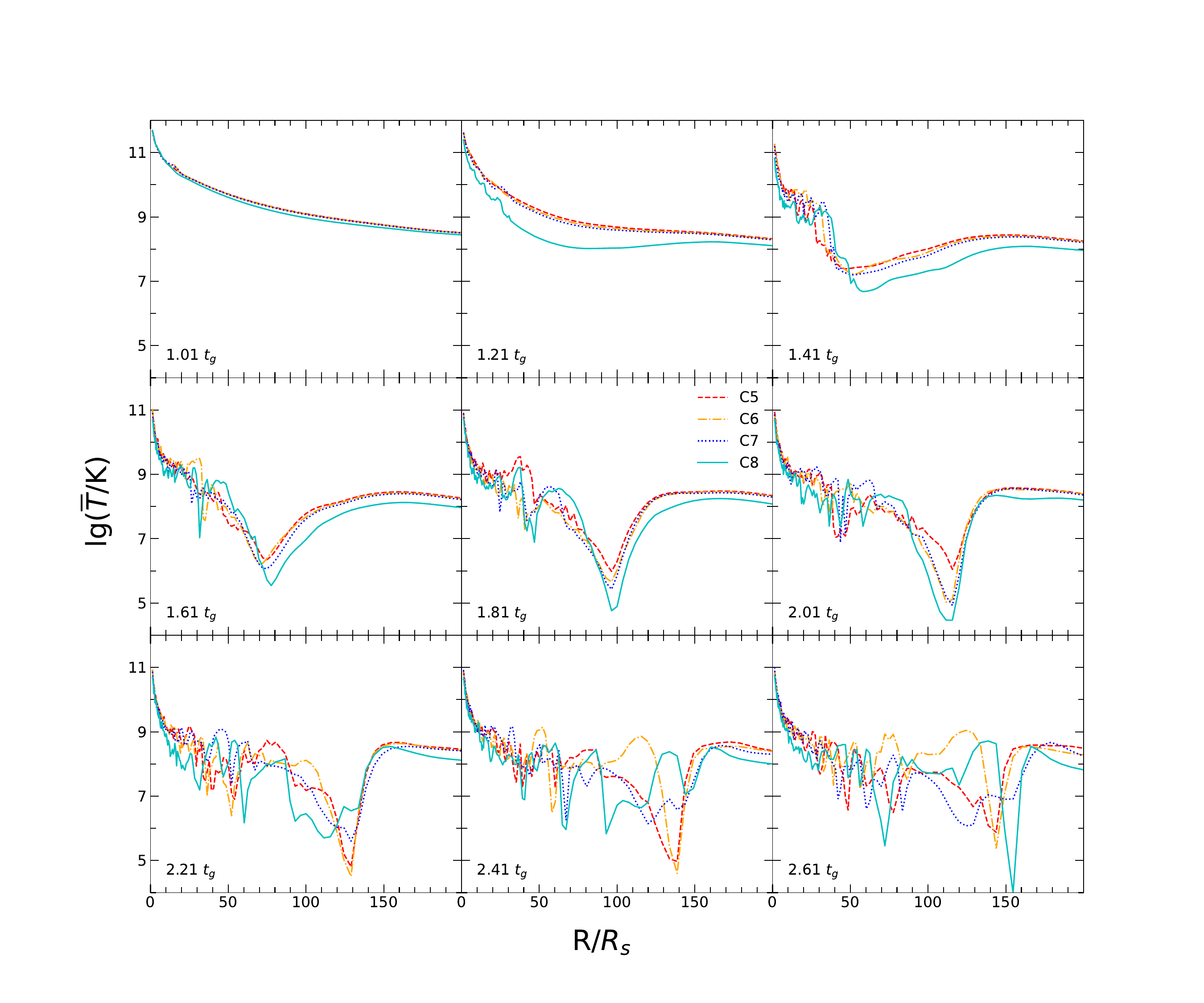}
\caption{\label{fig:develop}Radial distribution of mass averaged temperature at different evolutionary times for the models with cold clumps, i.e., Models C5-C8. The characteristic dips in the temperature curves indicate the positions of cold clumps.  } 
\end{figure*}

The radial distribution of the averaged temperature, or the distribution of cold clumps, are presented in Figure~\ref{fig:develop}.  
The red dashed, orange dash-dot, blue dotted and cyan solid lines correspond to Models C5 to C8, respectively.
The characteristic deep dips clearly indicate clump positions. And the temporal evolution of cold clumps can be found from the top left to the bottom right panels, of which the time intervals are all $0.2t_g$. 

All models exhibit similar evolutionary trends, differing primarily in magnitude and timescale. At $t=1.0t_g$, the temperature distributions are smooth and almost the same for different models. Subsequent radiative cooling progressively modifies the accretion flow, with the region around $50 R_S$ showing the most significant changes. After well-confined temperature dips form at $t=1.6t_g$, clumps begin migrating outward, reaching approximately $120 R_S$ by $t=2.0t_g$. In higher-accretion cases (Models C7 and C8), the original clumps fragment into smaller pieces, which subsequently migrate inward toward the black hole. Fragmentation occurs later in Models C5–C6. By $t=2.6t_g$, all models approach similar temperature distributions.
Similarly, there is another clump at around $30 R_S$, which moves outward to $50 R_S$ by $t=2.0t_g$ and then fragments.  

\begin{figure*}
\centering
\includegraphics[width=1\linewidth]{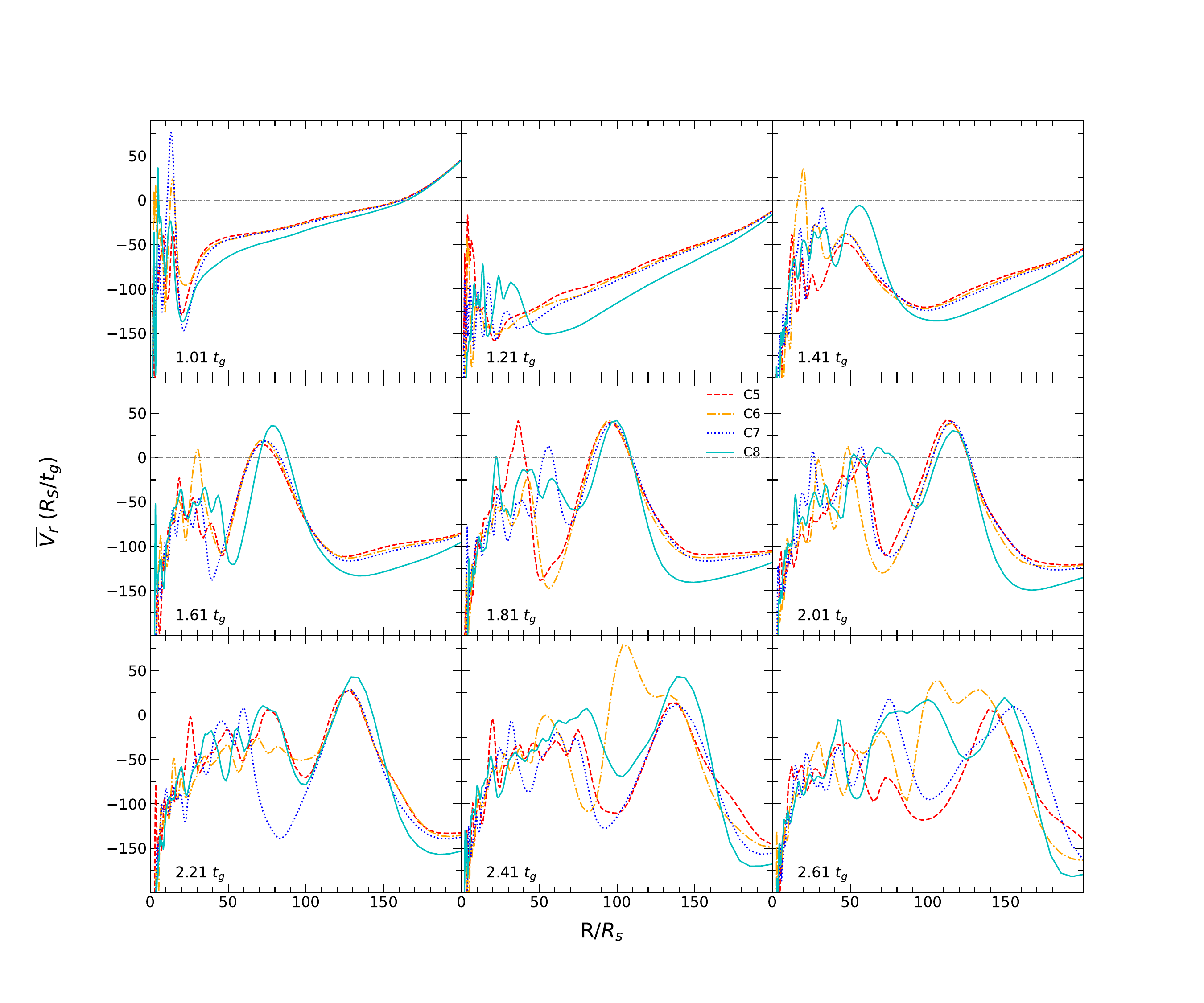}
\caption{\label{fig:vr} Radial distribution of mass averaged radial velocity at different evolutionary times for Models C5-C8. The line types are the same to Figure~\ref{fig:develop}. The gray dashed lines show the velocity of zero.} 
\end{figure*}

To further investigate the movement of the clumps, the mass weighted averaged radial velocity is plotted in Figure~\ref{fig:vr}. Consistent with Figure~\ref{fig:develop}, the radial velocity is positive at the position of temperature dips, which directly shows the clumps are moving outward. Moreover, it can also be seen that before the clump forms, some hot gas start to move outward when radiative cooling is introduced. This indicates that condense occurs together with outward movement. These condensed gas keeps cooling down and finally forms the clump when the temperature is lower than the defined critical temperature.

\subsection{Forces on the clumps}

The outward migration of the newly formed clumps during rising stage have never been reported in previous works, and therefore are worthy of further analysis.
Here we select Model C7 as example and concentrate on the two moments when the initial cold clump forms and when  the fragmentation of the clump begins. 
Since the movement of the clump is primarily along the radial direction, we focus on the radial forces. 
According to Equation \ref{Eq.2}, the forces include the gas pressure gradient, gravitational, viscous stress, and the centrifugal forces. 
Moreover, due to the simplified viscous stress we used, the radial component of the viscous stress force is zero. Consequently, the radial motion of the cold clumps is governed by three forces, i.e., the inward gravitational force ($F_g$), the outward centrifugal force (\textit{$F_r$}), and the pressure gradient force (\textit{$F_p$}). Notably, in the hot accretion flow, the gas pressure in the inner region is higher than that in the outer region, which means the pressure gradient force is also outward.
Therefore, the movement of the clumps can be inferred from the outward force $F_{out}\equiv F_r+F_p$, and the inward force $F_{in}\equiv F_g$.

\begin{figure}
\includegraphics[width=1\linewidth]{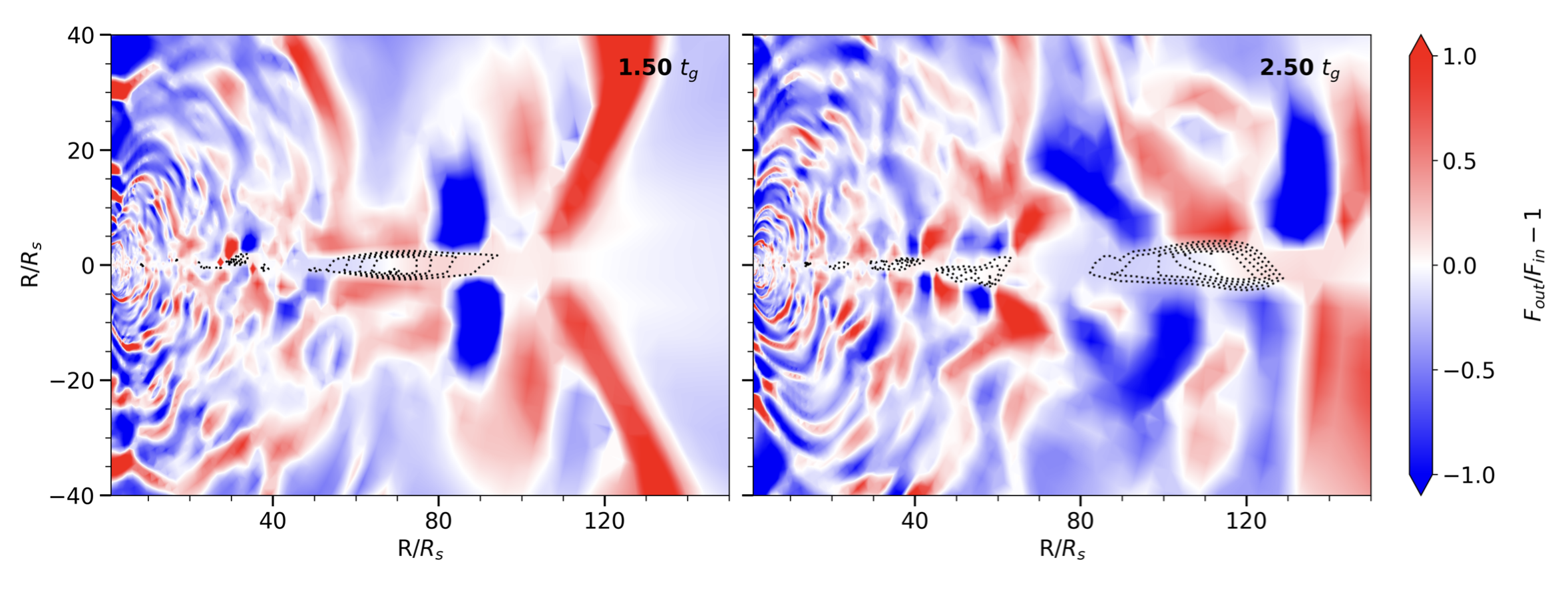}
\caption{\label{fig:force}The distribution of the ratio between outward and inward forces. We use model C7 because the phenomenon is more apparent than other model. The red and blue areas represent regions where the net force is directed outward or inward, respectively. The dotted lines are the contours of temperature, which shows the position and shape of the cold region.}
\end{figure}

The distribution of the ratio $(F_{out}-F_{in})/F_{in}$ is shown in Figure~\ref{fig:force}, where the positive and negative signs, or red and blue colors, mark the outward or inward direction of the net force, respectively.
It can be seen that the region of net inward or outward forces is fluctuating and mixing together. The region that cold clumps occupied shows significant variation of net force, indicating that the formation and evolution of the cold clumps occur in an unstable situation.

\begin{figure}
\includegraphics[width=1\linewidth]{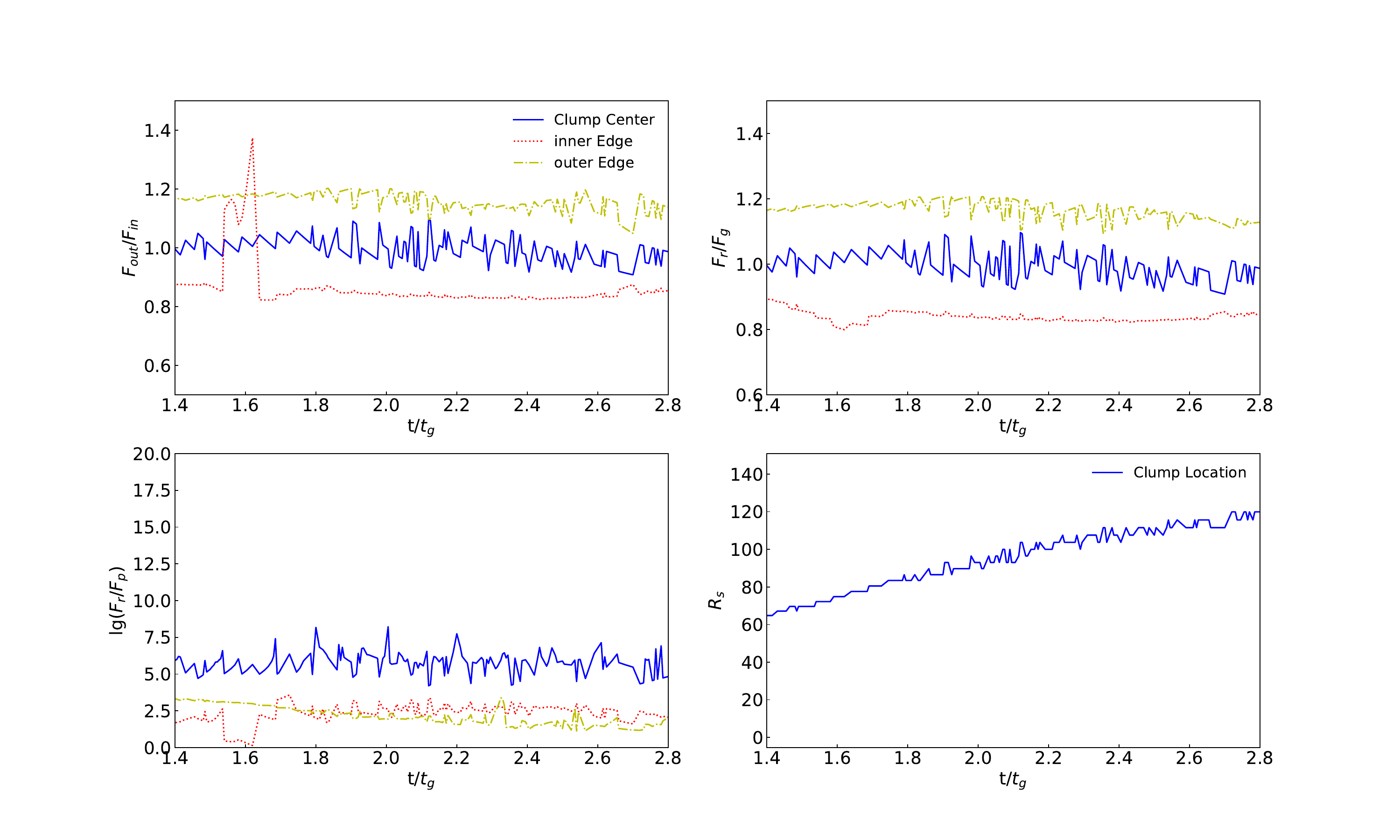}
\caption{\label{fig:forcedevelop}The relative contribution of different forces and the position of the clump center during $1.4-2.8~t_g$ of Model C7. The blue, red and yellow lines represents the central, inner and outer parts of the clump, respectively. 
The {\it{upper-left}} panel shows the ratio of the outward forces including centrifugal force $F_r$ and pressure gradient force $F_p$ to inward  gravitational force $F_g$. The {\it{upper-right}} panel shows the ratio of $F_r$ to $F_g$. The {\it{lower-left}} panel compares the two outward forces of $F_r$ and $F_p$, where logarithm is taken as the $F_p$ is much smaller than $F_r$. The {\it{lower-right}} panel shows the position of the clump center at different time.}
\end{figure}

To further check the forces on the clumps, we select three sample points on the clump, i.e., one point in the central region to represent the clump center, one point at the inner edge and one point at the outer edge, which move with the clump and remain the relative position. 
The variation of the forces on these three points during $1.4-2.8~t_g$ of Model C7 are shown in Figure~\ref{fig:forcedevelop}. 
First, it is very interesting to note that, 
at the clump center, the difference between the centrifugal force and gravitational force is small, although the time-averaged force ratio is slightly above one. This indicates the movement of the clump center is quasi-Keplerian.  At the outer and inner edges, centrifugal force is larger and smaller than gravity, and therefore the rotation of the gas are super- and sub-Keplerian, respectively.
The influences of gas pressure gradient should be ignorable because the gradient force is always two orders of magnitude smaller than gravitational force and centrifugal force as shown in the lower left panel.
The lower right panel shows the position of the clump center at different time. It is interesting to estimate the outward velocity of the clump, which is about $35 \, R_S/t_g$, or roughly about 35\% of the Keplerian velocity as $t_g$ corresponds to the orbital period at $100 R_S$.

During a period after the cold clump forming, the time-averaged centrifugal force at the clump center was slightly greater than the gravitational force. Consequently, the net force acting at the clumps was radially outward from the black hole over this period. Under this force regime, the center of the cold clumps began to move outward from the black hole. The center of the cold clumps moved from approximately $50 R_S$ to about $120 R_S$ in model C7.

It should be note that the net force at the inner edge of the clump is inward, or $F_{out}< F_{in}$, while the net force is outward at the outer edge, or $F_{out}> F_{in}$. The difference of the net forces should fragment the clump from the beginning. However, no obvious fragemtation are not found until the clump moves to $120 R_S$.
Why the clump does not fragment earlier should arise from two reasons. On the one hand, in spite of fragmentation, the net force at the inner boundary of the clump results in the splitting of small pieces of cold gas. Since the amount of cold gas splitting from the inner boundary of the clump is small, as well as its temperature is intermediate between the hot and cold gas, these part of gas can quickly change to hot gas without forming small clumps. On the other hand, during the rising stage, the mass of the clump increases with the increasing accretion rate, and the mass income reduces the effect of fragmentation. Therefore, the clumps do not show obvious fragment during the early stage.

After reaching the equilibrium position, the initial clump starts to fragment into small clumps in the inner part. 
We did not further check the force on the small clumps because of the limited spatial resolution and the fluctuation is serious.
Since these clumps are from the sub-Keplerian part of the original clump, they move inward after separating from the clump.

\subsection{Angular momentum of the clumps}

\begin{figure*}
\centering
\includegraphics[width=1\linewidth]{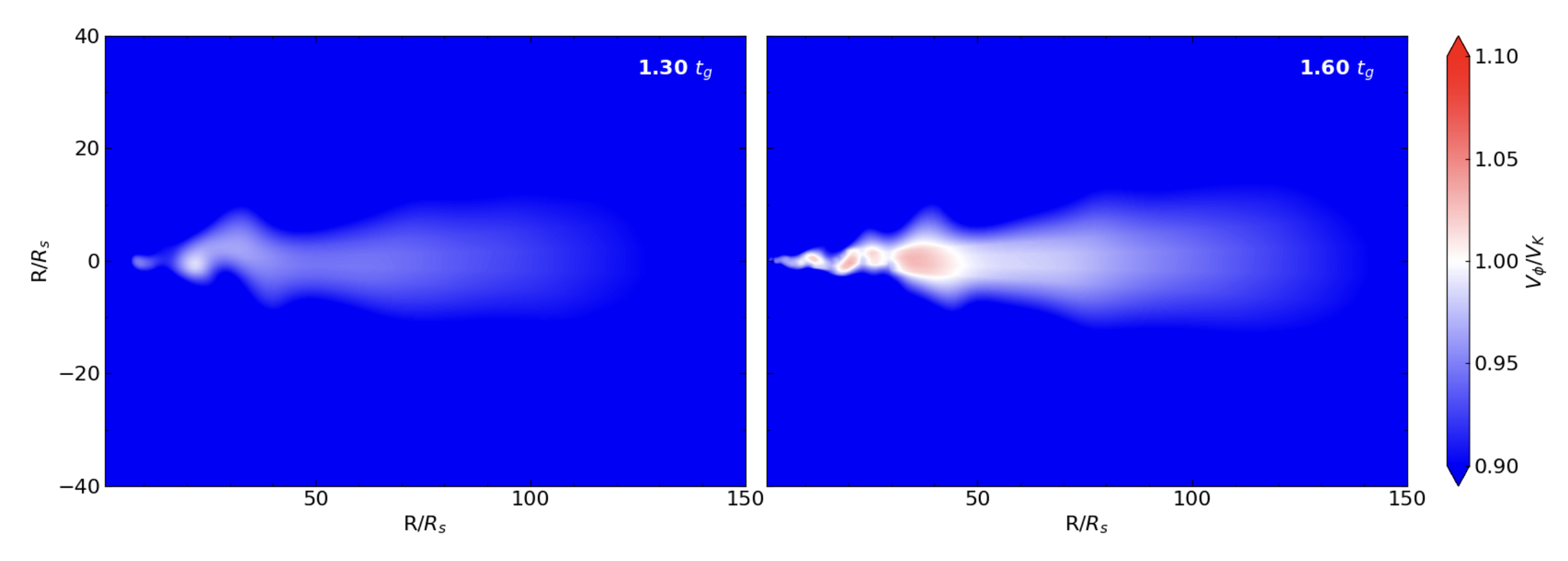}
\caption{\label{fig:v_clump} The toroidal velocity($V_\phi$) in hot accretion flow. This figure shows the toroidal velocity of Model C7, using a ratio method to compare the toroidal velocity at each grid point with the Keplerian velocity($V_K$). The red areas, where the ratio is greater than 1, indicate $v_\phi$ exceeding the $V_k$. It can be seen that the toroidal velocity at the clump exceeds the Keplerian velocity.} 
\end{figure*}

\begin{figure*}
\centering
\includegraphics[width=1\linewidth]{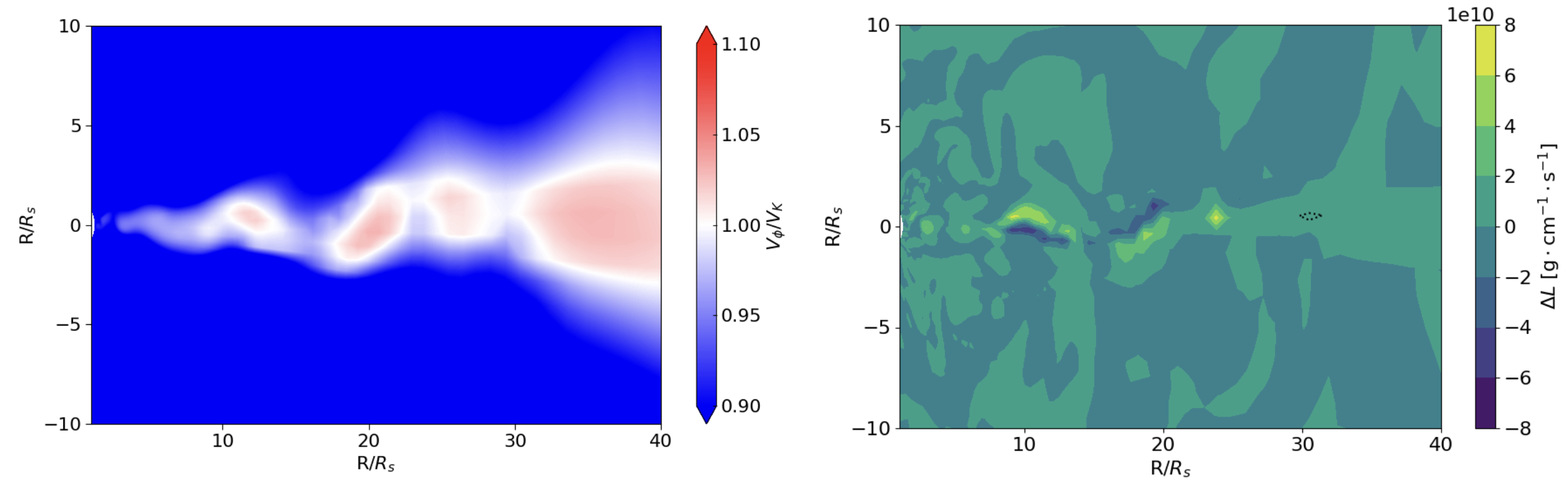}
\caption{\label{fig:deltaL}The toroidal velocity and change of angular momentum at time 1.60$t_g$ of Model C7. The left figure plots the toroidal velocity and the right plots the angular momentum change at the same moment. This figure represents the change in angular momentum compared to the previous moment, with negative values indicating a decrease in angular momentum and positive values indicating an increase at the net. It can be seen that there is a significant transfer of angular momentum near the clump.} 
\end{figure*}

Angular momentum provides another way to understand the evolution of clumps. 
In  Figure \ref{fig:v_clump}, the distribution of the azimuthal angular velocities of the accretion flow in Model C7 at two different moments are shown. We use the ratio of the azimuthal velocity to the Keplerian velocity as an indicator of the angular momentum. The ratio smaller/greater  than 1 describes the sub-\ /super- Keplerian case, and is marked with blue/red color in Figure.\ref{fig:v_clump}. 
At  $t=1.3t_g$, or before the clumps start to form,  the accretion flow is about Keplerian or sub-Keplerian.
When the clumps form, or $t=1.6t_g$, the super- and sub- Keplearian region are interspersed along the radial direction on the equatorial plane, and the clumps are super-Keplerian.
The alternative distribution of super- and sub- Keplerian regions shows that angular momentum are transferred between different radii.

There are two reasons for the redistribution of angular momentum.
One is the viscous torque between the inner and outer radii like normal disk.   
When a clump get more angular momentum, it suffers higher ram pressure from inflowing gas, its density increases  and consequently its temperature decreases further.
The other one is related with the condensing process. When the cold clumps form, the gas pressure that supporting the ambient gas becomes smaller, and the balance breaks. Consequently, the surrounding gas condense to the clump from larger or smaller radius. 
Because the accretion is proceeding, the angular momentum brought in by the gas from larger radii should be larger than that brought out by the gas from smaller radii, therefore the specific angular momentum of the clump becomes even larger, and the cold clumps become super-Keplerian.
The combination of these two reasons finally explain the outward moving of the clumps during rising stage.

To further track the transfer of angular momentum, Figure \ref{fig:deltaL} is plotted, in which the left panel is the same as the right panel of Figure \ref{fig:v_clump} but concentrates on smaller radial range, while the right panel shows the corresponding variation of the angular momentum.
It is interesting to find obvious transportation of angular momentum around the two clumps at $10R_S$ and $20R_S$.
Along the vertical direction, the angular momentum are transferred between the upper and lower layers, or along the radial direction, the angular momentum are conveyed from smaller to larger radii.
This feature can be easily understood with the above scenario that the condensing process is accompanied by the  increasing of angular momentum due to the gas from larger radii.
When the angular momentum are brought in, either by viscous torque or condensing, the clumps become super-Keplerian and moves outward, meanwhile the accretion process remains there, which implies the gas around the clumps should lose angular momentum and move inward. Therefore, at a given radius, the gas in clumps gains angular momentum and migrates outward. Conversely, the gas above or below the clumps reduces angular momentum and falls inward.

As the density of the clumps are higher than those in hot gas, the viscous force between the hot gas and clump are minor to the internal viscous force of the clumps. As a result, the clumps are similar to the isolated torus to some extent, the angular momentum being transferred from inner to outer region.
When the innermost part of the clump becomes sub-Keplerian, it separates from the rest of the clump and fall inward.
Therefore fragmentation takes place sequentially.

\section{Discussion}
In our simulation, the radiative cooling is dealt with in a rather simple way. The accretion flow is assumed to be optically thin and local emission rate is treated as the cooling rate. The radiation from the cold clumps are also ignored. Although such treatments are not very good, especially for the detailed process of state transition, such as the critical accretion rate and spectral evolution, they can simplify the simulation and tell us qualitatively how the clumps evolve.
Moreover, with more detailed processes or parameters, the critical conditions of clump formation should be more strict, which probably make it too difficult to find clumps in the simulation. 
As a preliminary work, we did not study the quantitative critical accretion rate when the clumps starts to form, but concerned on the qualitative results about the geometry and movement of the clumps. In this case,  our simplification does not affect the results seriously.

The movement of the central part of the clump is quasi-Keplerian.
This result agrees with the result of \citet{Wang2012}, in which the authors found that ,in the weak-coupling case, the azimuthal movement of the cold clumps can be described with Keplerian velocity.
In our simulation,  the viscous stress force inside the clump is much larger than that between the clump and outer hot gas because the viscosity adopted is simplified and depends only on the density, which agrees with the weak-coupling case.  
If the influence magnetic field, or other factors such as temperature, is considered, the difference between the internal and external viscous stress may be smaller, and the movement of clump center may deviate from Keplerian velocity, as predicted by \citet{Wang2012}. 
The influence of less simplified viscosity will be studied in future work.

Our simulations are 2-dimensional, in which axi-symmetric assumption is included. 
Whether the clumps take place in 3-dimensional simulation remain to be investigated.
If it remains true, condensing should happen not only along the poloidal plane but also along the azimuthal direction. 
On the one hand, the density of the clump could be even higher, which means the clump would be more like an isolated particle and move with Keplerian velocity; On the other hand, the fraction of the gas from larger radius should be smaller, which means the outward movement of the clump would be reduced. More detailed simulations will be conducted in future work.

In our simulation, we set ZEUS as a fluid model without the magnetic fields. In fact, different kinds and components of the magnetic fields may affect our results in different ways due to the freezing effect of plasma. 
Generally, the formation and outward moving of clumps should be depressed for the following reasons.
Firstly, the condensing of the hot gas is accompanied by the strengthening of the magnetic field, the pressure of which inversely inhibits condensation.
Secondly, no matter larger-scale or turbulent, the magnetic field makes the transport of angular momentum more efficient \citep{1982MNRAS.199..883B,yuan_numerical_2012}, and therefore the outward moving decreases.
Thirdly, the existence of magnetic field should induce the coupling between the hot gas and cold gas. As the hot gas is sub-Keplerian, it should reduce the angular momentum of the newly formed cold clumps.
So our results are proper for the cases of weak magnetic fields.

\section{Summary}

In this paper, based on hydrodynamic simulations of different parameters, we investigated the evolution of the cold clumps formed in the hot accretion flow. Agree with previous researches, the clumps appear when the accretion rate is higher than some critical value. 
In our simulations, more than one clumps is present initially.
The initially formed clumps may move outward due to excess of angular momentum that are transferred either by viscous torque from inner region or by condensing gas from larger radii. 
The inward moving of clumps is by fragmentation, with the inner part of the clumps losing angular momentum by internal viscosity. 
It is the first time that the influence of angular momentum on the evolution of clumps are discovered.

At present, our results about the geometric evolution of the clumps are just for the case of weak magnetic fields. Before further exploring the detailed process of state transition in 3-dimensional GRMHD simulations with radiative transfer \citep[e.g.,][]{Ohsuga2011,McKinney2014,Fragile2014,Ryan2015,Sadowski2017}, we are going to improve present study in 2-dimensional MHD simulations on the coupling between the hot gas and cold clumps \citep[e.g.,][]{2001ApJ...554L..49H,2003ApJ...592.1042I,yuan_numerical_2012}.

\section*{Acknowledgment}
This work is supported by the National Key R\&D Program of China (Grant No. 2023YFA1607902), the National SKA Program of China (No. 2025SKA0130100, 2020SKA0110100), and the NSFC (Grants 12192220, 12192223, 12347101, 12133008 and U2038108). X. Yang is  supported by Chongqing Natural Science Foundation (Grant CSTB2023NSCQ-MSX0093).

%\bibliography{sample631}{}

\begin{thebibliography}{}
\expandafter\ifx\csname natexlab\endcsname\relax\def\natexlab#1{#1}\fi
\providecommand{\url}[1]{\href{#1}{#1}}
\providecommand{\dodoi}[1]{doi:~\href{http://doi.org/#1}{\nolinkurl{#1}}}
\providecommand{\doeprint}[1]{\href{http://ascl.net/#1}{\nolinkurl{http://ascl.net/#1}}}
\providecommand{\doarXiv}[1]{\href{https://arxiv.org/abs/#1}{\nolinkurl{https://arxiv.org/abs/#1}}}

\bibitem[{Belloni(2010)}]{2010LNP...794...53B}
Belloni, T.~M. 2010, in Lecture Notes in Physics, Vol. 794, The Jet Paradigm, 53

\bibitem[{Blaes {et~al.}(2006)Blaes, Arras, \& Fragile}]{2006MNRAS.369.1235B}
Blaes, O.~M., Arras, P., \& Fragile, P.~C. 2006, MNRAS, 369, 1235

\bibitem[{Blandford \& Payne(1982)}]{1982MNRAS.199..883B}
Blandford, R.~D., \& Payne, D.~G. 1982, MNRAS, 199, 883

\bibitem[{Bu \& Gan(2018)}]{2018MNRAS.474.1206B}
Bu, D.~F., \& Gan, Z.~M. 2018, MNRAS, 474, 1206

\bibitem[{Bu {et~al.}(2019)Bu, Qiao, \& Yang}]{2019ApJ...875..147B}
Bu, D.~F., Qiao, E., \& Yang, X.~H. 2019, ApJ, 875, 147

\bibitem[{Dexter {et~al.}(2021)Dexter, Scepi, \& Begelman}]{2021ApJL...919...20D}
Dexter, J., Scepi, N., \& Begelman, M.~C. 2021, ApJL, 919, 20

\bibitem[{Esin {et~al.}(1997)Esin, McClintock, \& Narayan}]{Esin1997}
Esin, A.~A., McClintock, J.~E., \& Narayan, R. 1997, The Astrophysical Journal, 489, 865, \dodoi{10.1086/304829}

\bibitem[{{Fragile} {et~al.}(2014){Fragile}, {Olejar}, \& {Anninos}}]{Fragile2014}
{Fragile}, P.~C., {Olejar}, A., \& {Anninos}, P. 2014, \apj, 796, 22, \dodoi{10.1088/0004-637X/796/1/22}

\bibitem[{Hawley {et~al.}(2001)Hawley, Balbus, \& Stone}]{2001ApJ...554L..49H}
Hawley, J.~F., Balbus, S.~A., \& Stone, J.~M. 2001, ApJ, 554, L49

\bibitem[{Hayes {et~al.}(2006)Hayes, Norman, Fiedler, \& et~al.}]{2006ApJS..165..188H}
Hayes, J.~C., Norman, M.~L., Fiedler, R.~A., \& et~al. 2006, ApJS, 165, 188

\bibitem[{Igumenshchev {et~al.}(1996)Igumenshchev, Chen, \& Abramowicz}]{Igumenshchev1996}
Igumenshchev, I.~V., Chen, X., \& Abramowicz, M.~A. 1996, \mnras, 278, 236

\bibitem[{Igumenshchev {et~al.}(2003)Igumenshchev, Narayan, \& Abramowicz}]{2003ApJ...592.1042I}
Igumenshchev, I.~V., Narayan, R., \& Abramowicz, M.~A. 2003, ApJ, 592, 1042

\bibitem[{Kato {et~al.}(1998)Kato, Fukue, \& Mineshige}]{1998bhad.book.....K}
Kato, S., Fukue, J., \& Mineshige, S. 1998, Black-hole accretion disks (Kyoto Univ. Press)

\bibitem[{Liu \& Qiao(2022)}]{Liu2022}
Liu, B., \& Qiao, E. 2022, iScience, 25, 103544, \dodoi{https://doi.org/10.1016/j.isci.2021.103544}

\bibitem[{{Liu} {et~al.}(2011){Liu}, {Done}, \& {Taam}}]{Liu2011}
{Liu}, B.~F., {Done}, C., \& {Taam}, R.~E. 2011, \apj, 726, 10, \dodoi{10.1088/0004-637X/726/1/10}

\bibitem[{{Liu} {et~al.}(2007){Liu}, {Taam}, {Meyer-Hofmeister}, \& {Meyer}}]{Liu2007}
{Liu}, B.~F., {Taam}, R.~E., {Meyer-Hofmeister}, E., \& {Meyer}, F. 2007, \apj, 671, 695, \dodoi{10.1086/522619}

\bibitem[{{McKinney} {et~al.}(2014){McKinney}, {Tchekhovskoy}, {Sadowski}, \& {Narayan}}]{McKinney2014}
{McKinney}, J.~C., {Tchekhovskoy}, A., {Sadowski}, A., \& {Narayan}, R. 2014, \mnras, 441, 3177, \dodoi{10.1093/mnras/stu762}

\bibitem[{Mishra {et~al.}(2017)}]{2017MNRAS.467.4036M}
Mishra, B., {et~al.} 2017, MNRAS, 467, 4036

\bibitem[{Narayan \& Yi(1994)}]{1994ApJ...428...13N}
Narayan, R., \& Yi, I. 1994, ApJ, 428, 13

\bibitem[{{Ohsuga} \& {Mineshige}(2011)}]{Ohsuga2011}
{Ohsuga}, K., \& {Mineshige}, S. 2011, \apj, 736, 2, \dodoi{10.1088/0004-637X/736/1/2}

\bibitem[{Papaloizou \& Pringle(1984)}]{1984MNRAS.208..721P}
Papaloizou, J. C.~B., \& Pringle, J.~E. 1984, MNRAS, 208, 721

\bibitem[{Proga {et~al.}(2000)Proga, Stone, \& Kallman}]{2000ApJ...543..686P}
Proga, D., Stone, J.~M., \& Kallman, T.~R. 2000, ApJ, 543, 686

\bibitem[{Qiao \& Liu(2011)}]{Qiao2011}
Qiao, E., \& Liu, B.~F. 2011, The Astrophysical Journal, 744, 145, \dodoi{10.1088/0004-637x/744/2/145}

\bibitem[{Remillard \& McClintock(2006)}]{2006ARA&A..44...49R}
Remillard, R.~A., \& McClintock, J.~E. 2006, ARA\&A, 44, 49

\bibitem[{Rezzolla {et~al.}(2003{\natexlab{a}})Rezzolla, Yoshida, Maccarone, \& Zanotti}]{2003MNRAS.344L..37R}
Rezzolla, L., Yoshida, S., Maccarone, T.~J., \& Zanotti, O. 2003{\natexlab{a}}, MNRAS, 344, L37

\bibitem[{Rezzolla {et~al.}(2003{\natexlab{b}})Rezzolla, Yoshida, \& Zanotti}]{2003MNRAS.344..978R}
Rezzolla, L., Yoshida, S., \& Zanotti, O. 2003{\natexlab{b}}, MNRAS, 344, 978

\bibitem[{{Ryan} {et~al.}(2015){Ryan}, {Dolence}, \& {Gammie}}]{Ryan2015}
{Ryan}, B.~R., {Dolence}, J.~C., \& {Gammie}, C.~F. 2015, \apj, 807, 31, \dodoi{10.1088/0004-637X/807/1/31}

\bibitem[{Shakura \& Sunyaev(1973)}]{1973A&A....24..337S}
Shakura, N.~I., \& Sunyaev, R.~A. 1973, A\&A, 24, 337

\bibitem[{Shui {et~al.}(2023)Shui, Xie, Yan, \& Ma}]{Shui2023}
Shui, H.-Y., Xie, F.-G., Yan, Z., \& Ma, R.-Y. 2023, Research in Astronomy and Astrophysics, 23, 065020, \dodoi{10.1088/1674-4527/accdbe}

\bibitem[{Stone \& Pringle(2001)}]{Stone2011}
Stone, J., \& Pringle, J. 2001, Monthly Notices of the Royal Astronomical Society, 322, 461, \dodoi{10.1046/j.1365-8711.2001.04138.x}

\bibitem[{Stone {et~al.}(1999)Stone, Pringle, \& Begelman}]{1999MNRAS.310.1002S}
Stone, J.~M., Pringle, J.~E., \& Begelman, M.~C. 1999, MNRAS, 310, 1002

\bibitem[{Sadowski {et~al.}(2017)Sądowski, Wielgus, Narayan, Abarca, McKinney, \& Chael}]{Sadowski2017}
Sadowski, A., Wielgus, M., Narayan, R., {et~al.} 2017, Monthly Notices of the Royal Astronomical Society, 466, 705, \dodoi{10.1093/mnras/stw3116}

\bibitem[{Wang {et~al.}(2012)Wang, Cheng, \& Li}]{Wang2012}
Wang, J.-M., Cheng, C., \& Li, Y.-R. 2012, The Astrophysical Journal, 748, 147, \dodoi{10.1088/0004-637X/748/2/147}

\bibitem[{Wu {et~al.}(2016)Wu, Xie, Yuan, \& et~al.}]{2016MNRAS.459.1543W}
Wu, M.~C., Xie, F.~G., Yuan, Y.~F., \& et~al. 2016, MNRAS, 459, 1543

\bibitem[{Xie \& Yuan(2012)}]{Xie2012}
Xie, F.-G., \& Yuan, F. 2012, Monthly Notices of the Royal Astronomical Society, 427, 1580, \dodoi{10.1111/j.1365-2966.2012.22030.x}

\bibitem[{Xu {et~al.}(2020)Xu, Harrison, Tomsick, Walton, Barret, García, Hare, \& Parker}]{Xu2020}
Xu, Y., Harrison, F.~A., Tomsick, J.~A., {et~al.} 2020, The Astrophysical Journal, 893, 30, \dodoi{10.3847/1538-4357/ab7dc0}

\bibitem[{Yu {et~al.}(2018)Yu, Ma, Li, Zhang, \& Fang}]{Yu2018}
Yu, X.-D., Ma, R.-Y., Li, Y.-P., Zhang, H., \& Fang, T.-T. 2018, Monthly Notices of the Royal Astronomical Society, 476, 2045, \dodoi{10.1093/mnras/sty370}

\bibitem[{Yuan(2003)}]{2003ApJ...594L..99Y}
Yuan, F. 2003, ApJ, 594, 99

\bibitem[{Yuan {et~al.}(2012)Yuan, Bu, \& Wu}]{yuan_numerical_2012}
Yuan, F., Bu, D., \& Wu, M. 2012, The Astrophysical Journal, 761, 130, \dodoi{10.1088/0004-637X/761/2/130}

\bibitem[{Yuan \& Narayan(2014)}]{YN2014}
Yuan, F., \& Narayan, R. 2014, Annual Review of Astronomy and Astrophysics, 52, 529, \dodoi{https://doi.org/10.1146/annurev-astro-082812-141003}

\bibitem[{Zdziarski \& Gierliński(2004)}]{Zdziarski2004}
Zdziarski, A.~A., \& Gierliński, M. 2004, Progress of Theoretical Physics Supplement, 155, 99, \dodoi{10.1143/PTPS.155.99}

\end{thebibliography}
%\bibliographystyle{aasjournal}

\end{document}